\documentclass[aps,prd,twocolumn,showpacs,superscriptaddress,nofootinbib]{revtex4-1}

\usepackage[pdftex]{graphicx}

\usepackage{amsmath}
\usepackage[latin1]{inputenc}
\usepackage{txfonts}
\usepackage{hyperref}

\usepackage{xcolor}

\begin{document}

\title{Explosive synchronization with partial degree-frequency correlation}
\author{Rafael S. Pinto}
\email{rsoaresp@gmail.com}
\affiliation{Instituto de F\'\i sica ``Gleb Wataghin'', UNICAMP, 13083-859 Campinas, SP, Brazil.}
\author{Alberto Saa}
\email{asaa@ime.unicamp.br}
\affiliation{
Departamento de Matem\'atica Aplicada, 
 UNICAMP,  13083-859 Campinas, SP, Brazil.}

\date{\today}

\begin{abstract}
Networks of Kuramoto oscillators with a positive correlation between the oscillators frequencies and the degree of their corresponding vertices exhibits the so-called explosive synchronization behavior, which is now under intensive investigation. Here, we study and report explosive synchronization  in a situation that has not yet been considered, namely when only a part, typically small, of the vertices is subjected to a degree-frequency correlation. 
Our results show that in order to have explosive synchronization, it suffices to have degree-frequency correlations only for the hubs, the vertices with the highest degrees. 
Moreover, we show that a partial degree-frequency correlation does not only promotes but also allows explosive synchronization to happen in networks for which a full degree-frequency correlation would not allow it. 
We perform a mean-field analysis and our conclusions were corroborated by exhaustive numerical experiments for synthetic networks and also for the undirected and unweighed version of a typical benchmark biological network, namely  the neural network of the worm {\em Caenorhabditis elegans}. The latter is an explicit example where partial degree-frequency correlation leads to explosive synchronization with hysteresis, in contrast with the fully correlated case, for which no explosive synchronization is observed.
\end{abstract}

\maketitle

\section{Introduction}

Synchronization phenomena \cite{strogatz2003, pikovsky2003} manifest themselves in many and diverse areas. Some examples of current interest include  the biology of interacting fireflies \cite{buck1988}, cellular processes in populations of yeast \cite{monte2007}, audience clapping \cite{neda2000}, and power grids \cite{motter2013}, among many others.
Perhaps the most successful attempt to understand synchronization theoretically is the Kuramoto model \cite{kuramoto1975}. It has been heavily employed in the last decades as the paradigm to study the onset of synchronized behavior among nonidentical interacting agents, since it is one of the few models, together with some generalizations \cite{acebron2005}, that captures the essential mechanisms of synchronization and are still amenable to some analytical approaches  \cite{strogatz2000, ott2008}. 

The so-called Kuramoto model consists in an ensemble of $N$ oscillators, with phases and natural frequencies given, respectively, by $\theta_i$ and $\omega_i$,
placed on the vertices of a complex network \cite{arenas2008}. 
The network topology is described by the usual symmetric adjacency matrix $A_{ij}$, with elements $A_{ij} = 1$ if the vertices $i$ and $j$ are connected by an edge, and $A_{ij} = 0$ otherwise. The oscillators interact according to the equation
\begin{equation}
\frac{d\theta_i}{dt} = \omega_i + \lambda \sum_{j=1}^{N} A_{ij} \sin(\theta_j - \theta_i),
\label{kuramoto_definition}
\end{equation}
where $\lambda$ is the  coupling constant.
The global state of the oscillators (\ref{kuramoto_definition}) can be conveniently described by using the order parameter $r$
defined as
\begin{equation}
\label{r}
re^{i\psi} = \frac{1}{N}\sum_{j=1}^{N} e^{i\theta_j},
\end{equation}
which
 corresponds to the centroid of the phases if they are considered as a swarm of points moving around the unit circle. For incoherent motion, the phases are scattered on the circle homogeneously and $r \approx N^{-1/2}$ for large $N$, as a consequence of the central limit theorem, while for a synchronized state they should move in a single {lump} and,
 consequently, $r \approx 1$.
The general picture for the Kuramoto model is that, with very few exceptions, for small coupling strength $\lambda$ there is no synchronization and therefore $r \approx 0$ for large $N$. However, as one increases continuously the coupling constant $\lambda$, after passing a critical value $\lambda_{c}$, whose precise value depends both on the topology of the network and on the natural frequencies $\omega_i$ distribution, the order parameter $r$ starts to increase continuously. A sort of smooth second order phase transition from incoherence to synchronization takes place here.

Very recently, a new behavior for the Kuramoto model was discovered. 
In \cite{gomez-gardenes2011}, it was shown that in scale free networks, when there is a positive correlation between the natural frequencies of the oscillators and the degree of the vertices on which they lie, an abrupt first order transition from incoherence to synchronization, named explosive synchronization (ES), takes place. Typically, we also have a hysteresis behavior, and the forward and backward 
continuations ($r$ versus $\lambda$
diagram) do not coincide. 
In the simplest case exhibiting ES, the natural frequency $\omega_i$ of a given oscillator equals its vertex degree $k_i$,
\begin{equation}
\label{corr} 
\omega_i = k_i = \sum_{j=1}^{N} A_{ij}. 
\end{equation}
Explosive synchronization has also been observed in many other systems, as   the retarded Kuramoto model \cite{peron2012},  the second-order Kuramoto model \cite{ji2013}, in networks of FitzHugh-Nagumo oscillators \cite{chen2013}, and also
 in a network of chaotic Rösller oscillators \cite{levya2012}, allowing, in this case, an experimental observation of ES in electronic circuits. A mean-field approximation to explosive synchronization was applied in \cite{peron2012b}. We can also mention that a relation between explosive percolation \cite{achlioptas2009} and the generalized Kuramoto model proposed in \cite{zhang2013} was discussed in \cite{zhang2014}. We stress
 that there are other mechanisms capable of inducing first order phase transitions. For instance, in \cite{pazo2005}, an analytical treatment
 for first order phase transitions for synchronization is presented for the case of a Kuramoto model with 
 uniform distribution of the natural frequencies. The situation corresponding to ES is different, the frequencies are not randomly distributed,
 but subjected to the restriction (\ref{corr}).

Many works have recently been devoted to understand and to generalize the occurrence of explosive synchronization to other settings as, for instance, for weighted networks \cite{zhang2013, levya2013a}, where the coupling constant is no longer the same for all vertices, but its value varies for each pair of connected oscillators and may depend on the values of   their natural frequencies. In \cite{levya2013b}, starting from a given natural frequencies distribution, an algorithm was described to construct a network of oscillator exhibiting  ES.
However, in all these cases,   rather strong conditions to obtain ES are assumed. A first step to overcome this limitation was proposed in \cite{skardal2014}, where its shown that the addition of a quenched disorder to the degree-frequency correlation not only could maintain the ES, but could also 
induce ES in some kinds of networks without heterogeneous degree distributions. 
 
In this paper, we take another route and investigate ES in a Kuramoto model where \emph{only a few} of the vertices have a degree-frequency correlation.
We notice that the problem of partial correlation was briefly analyzed in \cite{gomez-gardenes2011} for the case of random correlations. They have shown that for a scale free network with exponent $\gamma = 2.4$, no ES was seen when less than around $50\%$ of the vertices had degree-frequency correlation. By means of a mean-field analysis, corroborated by exhaustive numerical experiments, we show that, in order to have ES, it suffices  that the degree-frequency correlation holds only for the hubs, the vertices with highest degree. We have found ES, for instance, in
Barabasi-Albert networks with only  $10\%$ of the vertices subjected to  degree-frequency correlation. More interestingly, we show that by restricting the degree-frequency correlation to the hubs  does not only promotes ES, but also allows it to happen in networks where the full degree-frequency correlation would not allow it. As we will see, this is the case, for instance, of a
typical benchmark biological network in the field:  the neural network of the worm {\em Caenorhabditis elegans}.

\section{A mean-field approach}

We will follow here the approach employed, for instance, in \cite{Ichi}.
For our networks, only vertices with degree $k$ larger than a threshold $k_*$ exhibit
the degree-frequency correlation (\ref{corr}), 
whereas the other vertices have random  natural frequencies with distribution $g(\omega)$. For these cases, the
corresponding  joint probability distribution for a vertex
with degree $k$ and natural frequency $\omega$ is given by
\begin{equation}
\label{defG}
G( \omega, k) = \left[\delta(\omega - k)P(k) - g(w)P(k)\right]H(k-k_*) + g(w)P(k),
\end{equation}
where $\delta(x)$, $H(x)$, and $P(k)$ are, respectively, the Dirac delta and the   Heaviside step functions, and
the  
network degree distribution. Notice that
\begin{equation}
 \int d\omega\, G( \omega, k)     = P(k) 
\end{equation}
and
\begin{equation}
  \int  dk \, G( \omega, k)  =
 P(\omega)H(\omega -k_*) + \alpha g(w),
\end{equation}
where
\begin{equation}
\alpha = \int_{k_{\rm min}}^{k_*}P(k) \, dk,
\end{equation}
with
$k_{\rm min} $ standing for 
  the network minimal degree. Furthermore, the network averages degree and frequency are given, respectively, by
\begin{eqnarray}
\langle k \rangle = \int  dk  \, k\int d\omega  G(\omega, k) = \int_{k_{\rm min}}^\infty kP(k)\, dk
\end{eqnarray} 
and
\begin{equation}
\label{defOmega}
\Omega = \int dk \int  d\omega \, \omega G(\omega, k) = \int_{k_*}^\infty kP(k)\, dk + \alpha  \langle \omega \rangle,
\end{equation}
where
\begin{equation}
\langle \omega \rangle =\int_{-\infty}^\infty \omega g(\omega)\, d\omega.
\end{equation}

Let us now consider the usual mean field  \cite{Ichi}  distribution density of oscillators $\rho(k,\omega,\theta,t)$ of vertices 
with phase $\theta$ at a time $t$, for given values of
the degree $k$ and frequency $\omega$, which is assumed to be normalized as
\begin{equation}
\int_0^{2\pi} \rho(k,\omega;\theta,t) \, d\theta = 1.
\end{equation} 
The probability $\cal P$ of a randomly chosen edge be attached to a degree $k$ vertex with phase
$\theta$ and frequency $\omega$ at time $t$  is given by
\begin{equation}
\label{prob}
{\cal P} = \frac{kG(\omega,k)\rho(k,\omega;\theta,t)}{\langle k \rangle}.
\end{equation}
 The usual
mean-field limit \cite{Ichi} for the Kuramoto network consists in employing (\ref{prob}) in the approximation of the right handed side of 
(\ref{kuramoto_definition}) for the description of the network average phase $\theta(t)$ 
\begin{eqnarray}
\label{conti}
\frac{d\theta}{dt} = \omega   
+ \frac{\lambda k}{\langle k \rangle} \int d\omega' & &  \int dk'\,k' G(\omega',k')  \times \\ \nonumber 
& &
\int d\theta'
\rho(k',\omega' ;\theta',t)\sin(\theta-\theta'). 
\end{eqnarray}
We now introduce the order parameter 
\begin{equation}
\label{ph}
re^{i\psi(t)} = \frac{1}{\langle k \rangle} \int d\omega'  \int dk'\,k'G(\omega',k')  \int d\theta'
\rho(k',\omega' ;\theta',t)e^{i\theta' },
\end{equation}
which incidentally does not correspond exactly to the continuous version of (\ref{r}), but it is indeed the more convenient one
for a mean-field analysis, see \cite{Ichi}, for instance, for further details. 
Of course, the onset of synchronization can be detected by using any
of the order parameters. As in (\ref{r}), $r$ is assumed to be real. By inserting the definition (\ref{ph}) in (\ref{conti}) we have finally  the simple expression
 \begin{equation}
\frac{d\theta}{dt} = \omega  + \lambda kr \sin(\psi-\theta),
 \end{equation}
 which is the standard mean-field equation for the Kuramoto network. 
 A convenient choice for studying the synchronization regime in our network is $\psi(t)=\Omega t + \psi_0$, where $\Omega$ is the
 network average frequency given by
 (\ref{defOmega}) and $\psi_0$ is an arbitrary phase. By introducing  $\phi(t) = \theta(t)-\psi(t)$, one has
 \begin{equation}
 \frac{d\phi}{dt} = \omega - \Omega - \lambda kr \sin\phi.
\end{equation}  
In terms of the new average phase $\phi$, the distribution density of oscillators must obey the
continuity equation \cite{peron2012b,Ichi}
\begin{equation}
\frac{\partial }{\partial t} \rho(k,\omega ;\phi,t) + \frac{\partial }{\partial \phi} 
\left( \frac{d\phi}{dt}\rho(k,\omega ;\phi,t) \right) = 0,
\end{equation}
which stationary solution $\rho(k,\omega ;\phi)$ is given bye the usual expression
\begin{equation}
\label{station}
\rho(k,\omega; \phi)  =   \left\{
\begin{array}{ll}
\delta\left( \phi -\arcsin \frac{ \omega -\Omega}{\lambda kr}\right), & {\rm for\ } | \omega -\Omega| \le \lambda kr ,\\ 
\frac{C_1(k,\eta)}{| \omega  -\Omega- \lambda kr \sin\phi|}, & {\rm otherwise},
\end{array}
\right.
\end{equation}
where
\begin{equation}
C_1(k,\omega) = \frac{\sqrt{(\omega-\Omega)^2 - (\lambda kr)^2}}{2\pi},
\end{equation}
  is a normalization constant. 
From (\ref{ph}) and (\ref{station}), we have for the stationary regime 
\begin{eqnarray}
\label{integrals}
re^{i\psi_0} =\frac{1}{\langle k \rangle} & & \int_{k_{\rm min}}^{\infty} dk\, k \times \\
  & & \left[
\int_{\Omega+\lambda kr}^{\infty} d\omega \,   G(\omega,k) 
\int d\phi\, \frac{C_1(\omega,k)e^{i\phi }}{\omega  -\Omega - \lambda kr \sin\phi}
  \right.   \nonumber \\
 +  & & 
\int_{\Omega-\lambda kr}^{\Omega+\lambda kr} d\omega \,  G(\omega,k) 
    \exp\left(i \arcsin \frac{\omega - \Omega}{\lambda kr} \right) \nonumber \\
   +  & &  \left.
\int_{-\infty}^{\Omega-\lambda kr} d\omega \,  G(\omega,k) 
\int d\phi\, \frac{C_1(\omega,k)e^{i\phi }}{\Omega  -\omega + \lambda kr \sin\phi}
  \right] .\nonumber
\end{eqnarray}
The first and third integral can be combined in an imaginary term we call $i\lambda rI_1(\lambda r)$, while the second one gives
origin to the real function $\lambda rI_2(\lambda r)$. From (\ref{integrals}), we have
\begin{equation}
\label{meanfield}
\langle k \rangle^2 = \lambda^2 \left[\left(I_1(\lambda r) \right)^2 + \left(I_2(\lambda r) \right)^2\right]
\end{equation}
for $r\ne 0$. The calculation details for $I_1$ and $I_2$ are presented in the Appendix. The corresponding
mean-field approximation \cite{Ichi} for the critical coupling $\lambda_c$ in   arises from the limit $r\to 0^+$ of equation
(\ref{meanfield})
\begin{equation}
\label{lambdalimit}
\lambda_c^2 = \lim_{r\to 0^+} \frac{\langle k \rangle^2}{\left(I_1(\lambda r) \right)^2 + \left(I_2(\lambda r) \right)^2},
\end{equation}
where 
\begin{eqnarray}
\label{ii}
\lim_{ r\to 0^+ }I_1(\lambda r) = & & {\rm p.v.\,} \left(\frac{1}{2}  \int_{k_*}^{\infty} dk\, \frac{k^2 P(k)}{k-\Omega}\right)  \\
& &\quad\quad \quad + {\rm p.v.\,} \left(\frac{ \beta}{2} \int_{-\infty}^{\infty} d\omega \frac{g(\omega)}{\omega - \Omega}\right) \nonumber
\end{eqnarray}
where p.v. stands to the Cauchy principal value for the integrals, with
\begin{equation}
\label{beta}
\beta = \int_{k_{\rm min}}^{k_*}k^2 P(k) \, dk,
\end{equation}
and 
\begin{equation}
\label{limit1}
\lim_{r\to 0^+} I_2(\lambda r) = \frac{\pi}{2}\left(\beta g(\Omega) + \Omega^2P(\Omega)H(\Omega - k_*)\right).
\end{equation}

There are several special cases we might consider now in order to test the predictions of our mean-field analysis. 
For a Barabàsi-Albert (BA) network ($P(k) \propto k^{-3}$), for instance, we would have
\begin{equation}
\label{BA}
{\rm p.v.\,} \left(   \int_{k_*}^{\infty} dk\, \frac{k^2 P(k)}{k-\Omega}\right)  = \frac{2k_{\rm min}^2}{ \Omega}\log \left|\frac{k_*}{ k_*-\Omega}\right|.
\end{equation}
If we assume now a  symmetrical $g$ around $\Omega$, {\em i.e.},   $g(\Omega + \omega) = g(\Omega - \omega)$, 
implying of course that   $\Omega = \langle \omega \rangle$, we have    from  (\ref{defOmega}) 
\begin{equation}
\Omega = \frac{\int_{k_*}^\infty kP(k)\, dk}{\int_{k_*}^\infty  P(k)\, dk} =2 k_*.
\end{equation}
For this case, both integrals in (\ref{ii}) vanish,   
leading to the following critical coupling $\lambda_c$ for a BA network with $\Omega = \langle \omega \rangle$
\begin{equation}
\lambda_c = \frac{2\langle k \rangle}{\pi\left(\beta g(\Omega) + \Omega^2P(\Omega)\right) }.
\end{equation}
The other cases we will consider here are those ones with vanishing $\langle \omega \rangle$. For these cases, we have typically $\Omega < k_*$. We can evaluate easily the second integral in (\ref{ii}), for instance, in the case of a  homogeneous $g(\omega)$ with null average and compact support, {\em i.e.}, for
\begin{equation}
\label{homo}
g(\omega) = \left\{
\begin{array}{ll}
\sigma_0^{-1} & {\rm for\ }|\omega| \le \frac{\sigma_0}{2},\\
0 & {\rm otherwise}.
\end{array}
\right.
\end{equation}
In this case, we have for a BA network with $\Omega < k_*$  
\begin{equation}
\lambda_c^2 = \frac{4\langle k \rangle^2}{ \pi^2\beta^2 g(\Omega)^2   +
\left(\frac{2k_{\rm min}^2}{\Omega}\log \frac{k_*}{  k_*-\Omega } + \frac{\beta}{\sigma_0}\log\left|\frac{\sigma_0-2\Omega}{\sigma_0+2\Omega}\right|\right)^2}.
\end{equation}
On the other hand, for a standard Gaussian distribution
\begin{equation}
\label{gaussian}
g(\omega) = \frac{1}{\sigma_0\sqrt{2\pi}}\exp\left( - \frac{\left(\omega-\langle \omega\rangle\right)^2}{2\sigma_0^2} \right)
\end{equation}
we have (see the Appendix for the calculation details)
\begin{eqnarray}
\label{pvgaussian}
{\rm p.v.\,} \left( \int_{-\infty}^{\infty} d\omega \frac{g(\omega)}{\omega - \Omega}\right) = & & 
\frac{1}{\sigma_0}\sqrt{\frac{\pi}{2}} {\rm erfi}\left(\frac{\langle \omega\rangle-\Omega}{\sqrt{2}\sigma_0} \right)\times \nonumber  \\
& & \quad\quad \exp\left( - \frac{\left(\langle \omega\rangle-\Omega\right)^2}{2\sigma_0^2} \right),
\end{eqnarray}
giving origin  consequently to another expression for $\lambda_c$ in the mean-field approximation. 
 Notice that the first integral in (\ref{ii}) cannot be evaluated in general in term of elementary 
functions as it was done for BA networks. Generic power laws  degree distributions $P(k)\propto k^{-\lambda}$, with real $\lambda > 2$, for instance, are examples of cases where the integral cannot be evaluated in closed form. However, a series solution is indeed available, see the Appendix.
For $P(k)\propto k^{-n}$ with integer $n>2$, we have
\begin{eqnarray}
\label{intn}
{\rm p.v.\,} \left( \int_{k_*}^{\infty} dk  \frac{k^2P(k)}{k-\Omega}\right) = & & (n-1)k_*
\left(\frac{k_*}{\Omega}\right)^{n-2}\left(\log \left| \frac{k_*}{k_*-\Omega}\right| \right. \nonumber \\
& & \left. 
-  \sum_{\ell=1}^{n-3} 
\frac{1}{\ell} \left(\frac{\Omega}{k_*}\right)^{\ell} \right)  ,
\end{eqnarray}
 For a exponential
distribution $(P(k)\propto e^{-\gamma k})$ , on the other hand, it is also possible to evaluate the integral exactly, leading to 
\begin{equation}
\label{pvexp}
{\rm p.v.\,} \left( \int_{k_*}^{\infty} dk \frac{k^2P(k)}{k-\Omega}\right) =  \frac{k_*+1}{\gamma} +\Omega -
 \gamma \Omega^2 e^{-\gamma\Omega+k_*}{\rm Ei}(\gamma(\Omega-k_*))   ,
\end{equation}
where
\begin{equation}
{\rm Ei}(x) = - \int_{-x}^\infty \frac{e^{-t}}{t}\, dt
\end{equation}
is the standard exponential integral function. 

Several important conclusions arises from our predicted value of $\lambda_c$. For instance, consider the case $k_*\to \infty$ or, in other words,
the case without any degree-frequency correlation. Let us also assume $g(\omega) = g(-\omega)$ and, hence $\Omega = \langle \omega\rangle = 0$.
In this case, we recover the usual result \cite{Ichi}
\begin{equation}
\lambda_c = \frac{2\langle k\rangle }{\pi\beta g(0)},
\end{equation}
with $\beta = \langle k^2\rangle$. For a network such that $\beta = \langle k^2\rangle\to \infty$ (this is the case, for
instance, of BA networks), we would have the well known result $\lambda_c\to 0$, meaning that no phase transition should be present at all, {\em i.e.}, synchronization should
appear continuously as $\lambda_c$ increases starting from zero.  This is a case where we should  expect neither explosive synchronization nor
second order transitions.
Let us now consider in this same network a partial degree-frequency correlation, {\em i.e.}, let us consider the case of finite $k_*$. Notice that
$\beta$ now is finite. In fact, with only one possible exception, all the terms contributing to the denominator of $\lambda_c$ in
(\ref{lambdalimit}) will be finite in this case, implying that $\lambda_c>0$, {\em i.e.}, there must exist a sudden transition from $r=0$ (incoherence) 
to $r\ne 0$ (synchronization). In other words, a partial degree-frequency correlation suffices to induce a phase transition
 in this network.
As we will see, this transition can be an explosive synchronization or  a second order phase transition, depending on the value of
$\Omega$.
  The exception quoted above
corresponds to the case where $\Omega=k_*$, which from (\ref{BA}) implies in $I_1(0^+)\to\infty$ and, hence,   the suppression of ES. See the Appendix for further details.

We need also to comment the case $k_*=k_{\rm min}$, {\em i.e.}, the case of total correlation considered, for instance, in \cite{peron2012b}.
For this case, $\beta =0$ and all references to $g(\omega)$ in the critical coupling expressions disappear, as it is indeed expected, and we have, for a BA network,
\begin{equation}
\lambda_c  = \frac{2\langle k \rangle }{\pi   \langle k \rangle^2P(\langle k \rangle) },
\end{equation}
since $\Omega = \langle k\rangle   = 2k_*$. 
This is the expression obtained in  \cite{peron2012b}. However, it is valid only for BA networks, for which (\ref{BA}) vanishes. For any other
degree distribution function, one needs to include the term corresponding to $I_1(0^+)$. This extra term is   absent in the
analysis of \cite{peron2012b}. Anyway, it does not alter the prediction of finite $\lambda_c$, {\em i.e.}, the presence
of a phase transition in the fully correlated case for any degree distribution function. For a fully correlated network with a degree distribution $P(k)\propto k^{-n}$,  $n>2$, we have
\begin{equation}
\Omega = \langle k \rangle = \frac{n-1}{n-2}k_{\rm min},
\end{equation}
and the correct expression for the
critical coupling $\lambda_c$ is
\begin{equation}
\lambda_c = 
 \frac{2\frac{(n-1)^{n-2}}{(n-2)^{n-1}} }{\sqrt{\pi^2  +  \left(\log (n-2) 
- \sum_{\ell = 1}^{n-3}\frac{1}{\ell}\left(\frac{n-1}{n-2}\right)^\ell
 \right)^2}},
\end{equation}
valid for any integer $n>2$.

\section{Numerical Results}

We have performed exhaustive numerical experiments not only to test our mean-field analysis, but mainly to gain some
knowledge  in   situations for which the mean-field approach cannot be directly employed.  This is  the case, for instance, of
networks  which we do not know a priori the degree distribution $P(k)$. Another situation is the discerning of first and second
order phase transitions, as we will see below. In general, the process of synchronization can be numerically analyzed by computing the forward and backward synchronization diagrams $r(\lambda)$ 
according to Ref. \cite{gomez-gardenes2011}. The forward continuation is performed by starting with an initial value $\lambda_0$ of the coupling constant.  We numerically solve  equations (\ref{kuramoto_definition}) with random initial conditions
for $\lambda=\lambda_0$ and evaluate the  order parameter $r(\lambda)$
in the stationary regime. Then we increase the coupling by a small value $\delta \lambda$ and, using the outcome of the last run as the initial condition, calculate the new value of the stationary order parameter $r(\lambda+\delta \lambda)$. We repeat these steps until a maximal value 
 $\lambda_1$ is reached. In the same way, the backward continuation is done by decreasing the coupling by steps of size $\delta \lambda$ from the maximal value of $\lambda_1$. In all of the results presented here, we used $\delta \lambda = 0.02$, but our conclusions do
 not depend on the value of the increment.
We  also compute how the oscillator effective frequencies $\Omega_i$, defined as
\begin{equation}
\Omega_i = \frac{1}{T} \int_{\tau}^{\tau + T} \dot{\theta}_{i}(t) dt,
\label{effective_freq}
\end{equation}
vary as function of the coupling constant $\lambda$. 
In all of the numerical experiments performed in this work, both $r(\lambda)$ and $\Omega_i$ were evaluated by solving the system 
 (\ref{kuramoto_definition}) up to a time $\tau = 340$ time units. Then all quantities  were averaged over the next time interval of length $T = 110$ time units. Again, our results does not depend considerably on the choices of $\tau$ and $T$, provided that $\tau$
 is large enough to assure that the system is in a stationary regime and that $T$ is compatible with our statistical analysis.  
Our numerical computations were done by using the SciPy package for python \cite{SciPy}. The system of ordinary differential equations (\ref{kuramoto_definition}), in
particular, is solved with SciPy {\tt odeint} routine, which is indeed an implementation of {\tt lsoda} from the FORTRAN library odepack, and
it is known to be effective and efficient for stiff  system of ordinary differential equations. Since the
oscillator 
frequencies
$\omega_i$ can vary considerably over the network, the numerical integration of (\ref{kuramoto_definition}) must be
done cautiously.

\begin{figure}[t]
\centering
\includegraphics[scale=0.20]{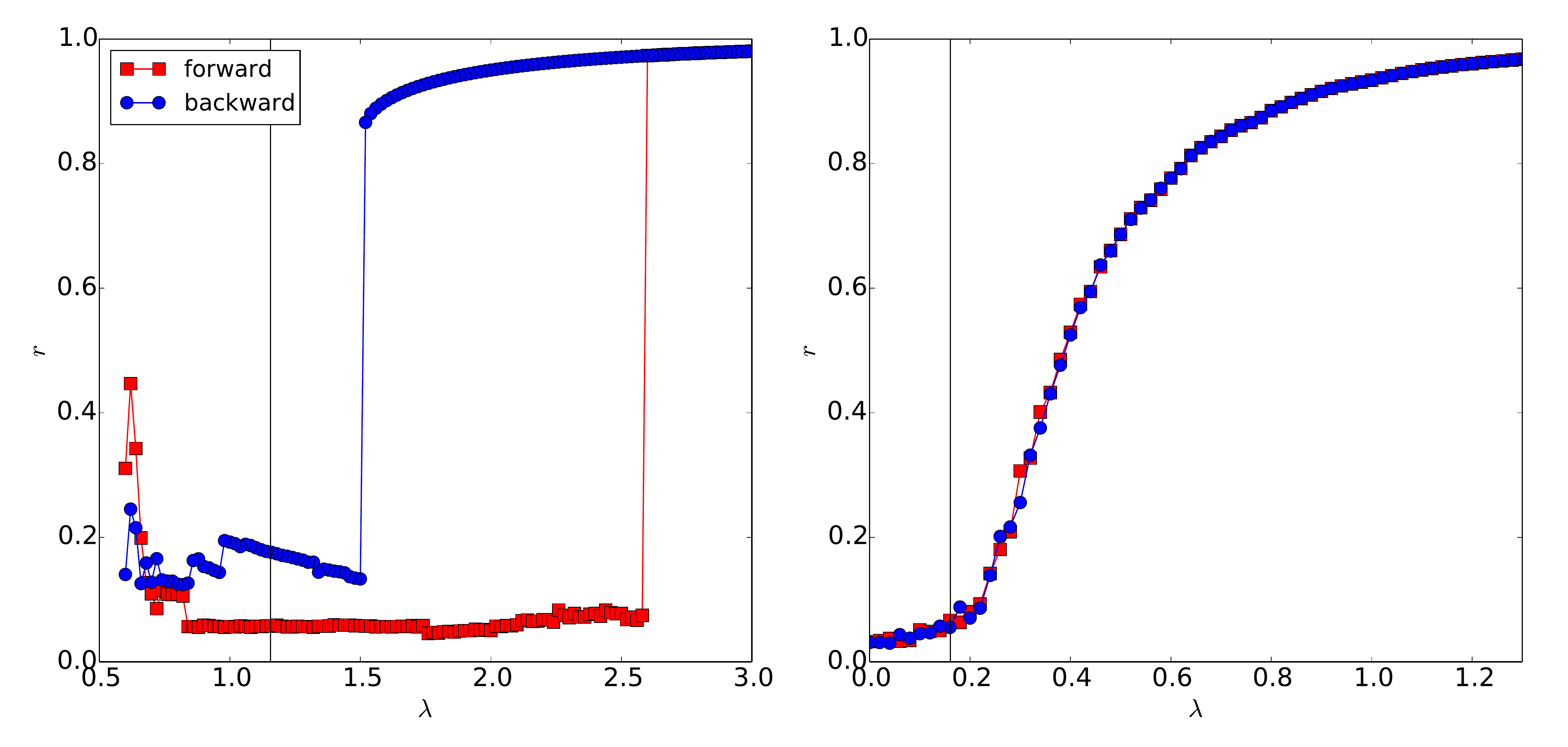}
\caption{Synchronization diagrams $r(\lambda)$ for a Barabasi-Albert network with $k_{\rm min}=3$ and
$\langle k\rangle = 6$, with $N=1000$ vertices, and $k_*=10$, corresponding approximately to only 10\% of vertices
with degree-frequency correlation. The left panel corresponds to a situation where the frequencies of the uncorrelated
vertices were drawn from the null average homogeneous probability distribution  (\ref{homo}) with $\sigma_0=1$, while
the right one corresponds to the Gaussian (\ref{gaussian}) with $\Omega$ given by (\ref{defOmega}) and $\sigma_0=1/2$. In both cases,
the mean-field analysis predicts $\lambda_c$ (the vertical line), but cannot discern between the explosive
synchronization (first order phase transition) of the left panel from the continuous second order phase transition of the
right panel. Our numerical analysis shows that, typically, greater values of $\Omega$ tend to 
favor second order phase transitions instead of ES. The simulations corresponding to the $\langle \omega \rangle = 0$ case 
 (left panel) were plagued by large statistical
fluctuations for small $\lambda$ which origin is unclear. For small $\lambda$, forward and backward continuation do not coincide  and large deviations are observed for different runs. Both panel depicts only one run to put in evidence these discrepancies.
}
\label{meanfieldfig}
\end{figure}

We perform many numerical simulations in order to test the predictions of the mean-field analysis of the last section.  Figure (\ref{meanfieldfig}) 
depicts a typical situation where  $\lambda_c$  is calculated for a explosive synchronization
case and for a second order phase transition.
 Although one can predict the occurrence  of phase transitions by evaluating the  critical coupling $\lambda_c$ 
in the mean-field approximation,   
one cannot advance  if the corresponding  transition is a continuous second order phase transition  or an explosive synchronization
phenomenon. 
Also, the mean-field analysis cannot predict the intensity, {\em i.e.}, the size of the hysteresis loop, for the case of ES. We use our numerical experiments not only to corroborate the mean-field analysis, but mainly to explore these points that are, in principle,
inaccessible analytically with the approach of the last section.  In our numerical experiments we will present here, 
we consider basically two kinds of networks. First, we analyze the existence of ES in synthetic networks constructed with the mechanism proposed in \cite{gomez-gardenes2006}, and later we will also study the existence of explosive synchronization in the neural network of the worm {\em Caenorhabditis elegans}.

\subsection{Synthetic Networks}

The synthetic networks considered here were constructed according to the mechanism introduced in \cite{gomez-gardenes2006}, which  depends only  upon the parameter $\alpha$, with $0 \leq \alpha \leq 1$. It is essentially  a growing mechanism where the newly added vertex attaches to a randomly chosen vertex with probability $\alpha$, or to higher degree vertices with probability $1 - \alpha$. In this way, by tuning a single parameter, we can build networks with varying heterogeneity, measured by the degree distribution $p(k)$. For $\alpha = 1$, we have Erd\H{o}s-R\'enyi  networks with a exponential decaying degree distribution, while for $\alpha = 0$ we have Barabàsi-Albert networks  with a power law degree distribution $p(k) \propto k^{-3}$. We have considered networks with $N = 500$ vertices and mean degree $\langle k \rangle = 6$, but our results do not depend on the networks details, provided they are sufficient to our statistical analyses.

\begin{figure*}
\centering
\includegraphics[scale=0.40]{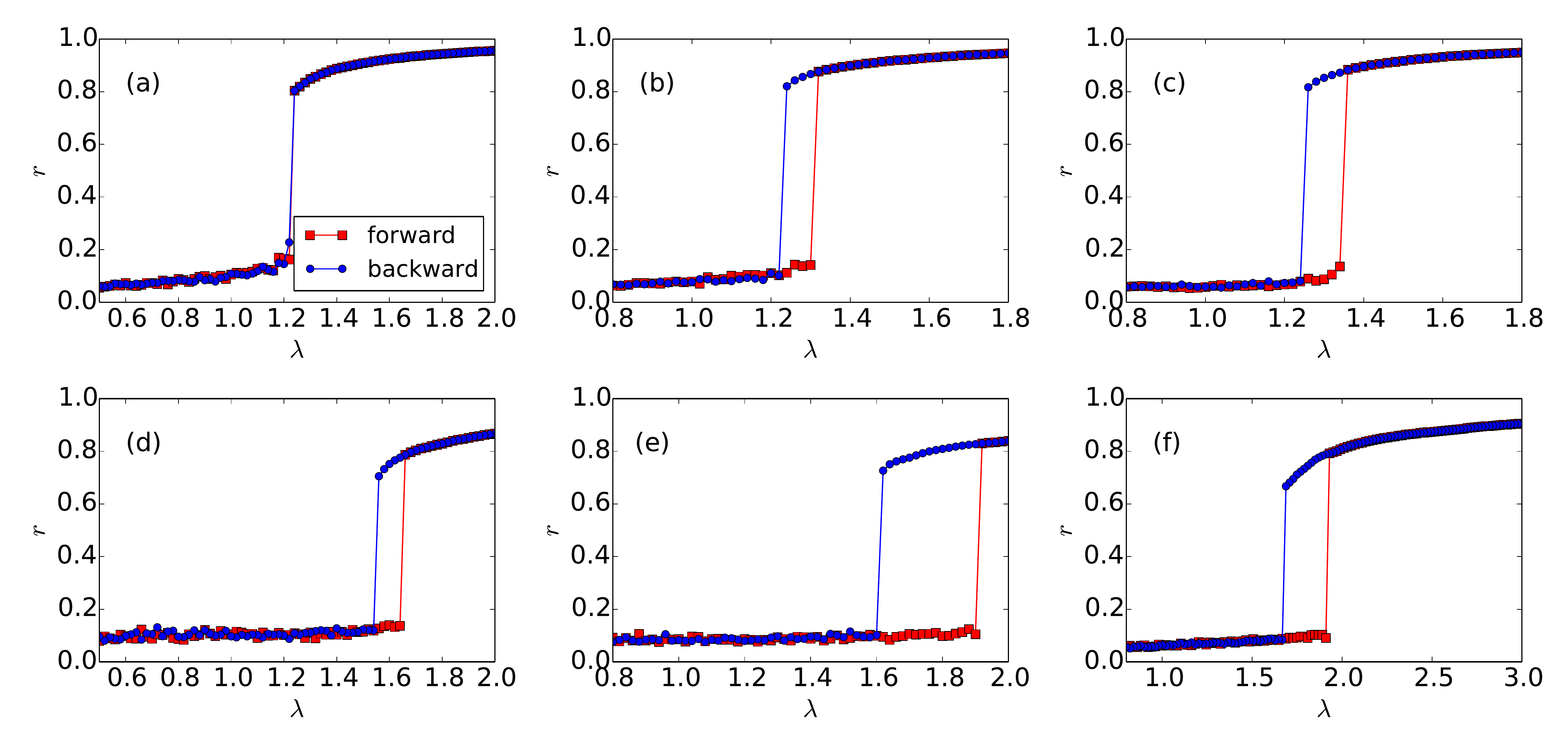}
\caption{The graphics show the synchronization diagrams $r(\lambda)$ for networks built with the mechanism proposed in \cite{gomez-gardenes2006}. The panels (a), (b), and (c) show, respectively, the forward and backward continuations for full degree-frequency correlation and $\alpha = 0.2$, $\alpha = 0.1$, and $\alpha = 0$. The bottom panels (d), (e) and (f) are the diagrams when only $10\%$ of the vertices with largest degree have degree-frequency correlation. See the text for further details.}
\label{ba_er_diagram}
\end{figure*}

The top panels (a), (b), and (c) of Figure \ref{ba_er_diagram} show the synchronization diagrams for networks with full degree-frequency correlation according to (\ref{corr}). The values of $\alpha$ are, respectively, $\alpha = 0.2$, $\alpha = 0.1$ and $\alpha = 0$. On the other hand, the bottom panels, (d), (e), and (f) depict synchronizations diagrams for the same networks, but now having degree-frequency correlation only for 50 largest degree vertices, while for the remaining ones their natural frequencies $\omega_i$ were drawn from a power law distribution $g(\omega) \propto \omega^{-\gamma}$ with exponent $\gamma = 3$. The frequencies $\omega_i$ are obtained by using the standard inversion
method (see, for instance, \cite{NR}), {\em i.e.}, if $x$ is a random variable with uniform distribution on the interval $[0,1)$, then 
\begin{equation}
\omega = \frac{\omega_{0}}{(1 - x)^{1/(  \gamma-1)}}
\label{powerlaw_definition}
\end{equation}
is a random variable with a power law distribution $g(\omega) = \omega^{-\gamma}$ on $[\omega_0,\infty)$, with $\gamma>1$. We have used
$\omega_0=3$ in the simulations of Figure \ref{ba_er_diagram}.

\begin{figure}
\centering
\includegraphics[scale=0.20]{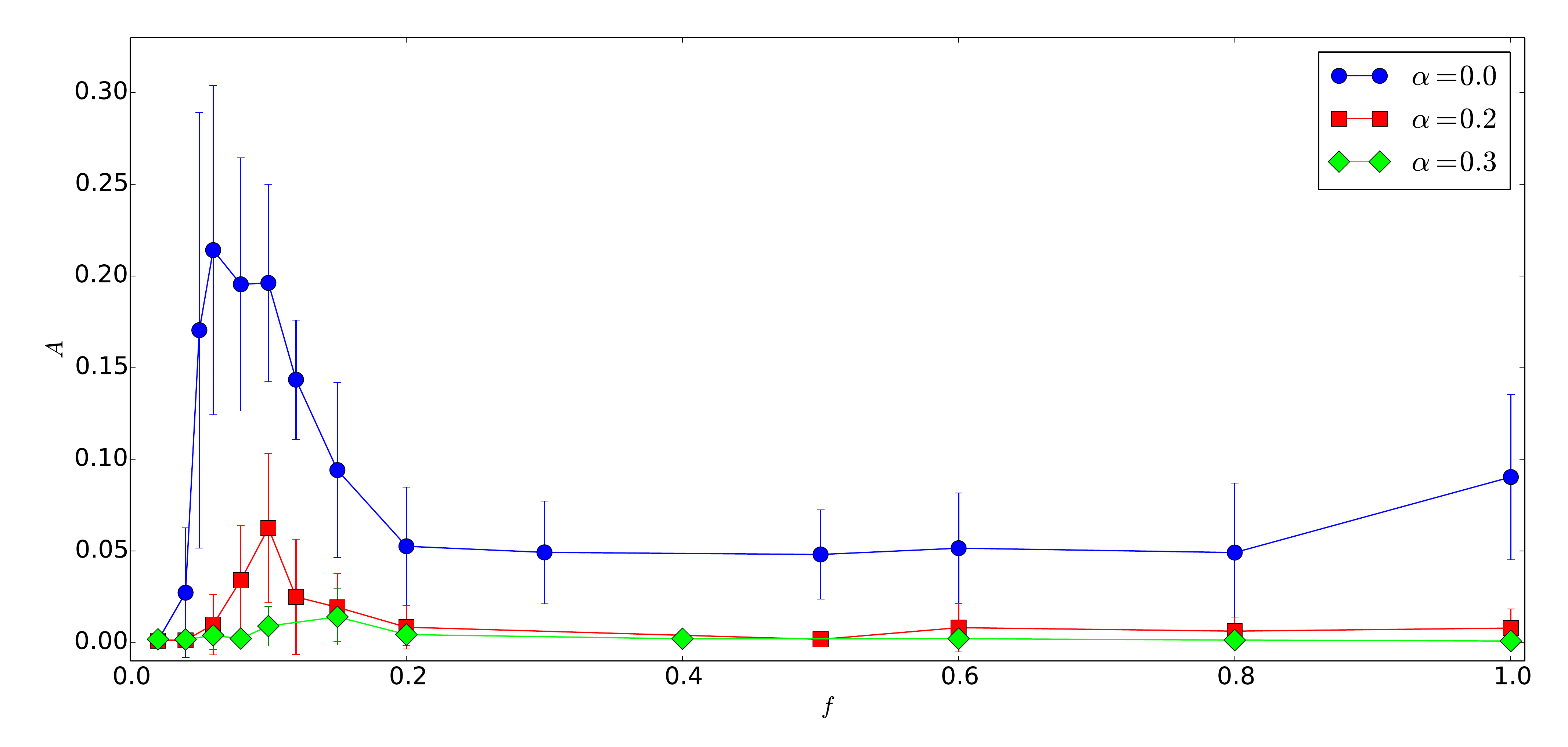}
\caption{The graphic shows the area $A$ between the forward and backward continuations of the synchronization diagrams $r(\lambda)$ as a function of the fraction $f$ of vertices for which degree-frequency correlation holds for networks built with the mechanism proposed in \cite{gomez-gardenes2006}.  Different curves show the behaviour of $A$ for different values of the parameter $\alpha$  that measures the heterogeneity of the degree distribution. Each point corresponds to an average over 10 different sets of networks and natural frequencies and the errorbars represent the corresponding standard deviations.}
\label{hist_area}
\end{figure}

It is interesting to notice that by imposing a partial degree-frequency correlation,   not only we  keep the explosive synchronization in the cases it already happens with full correlation, {\em i.e.}, panels (b)-(e) and (c)-(f), but somehow unexpectedly, ES emerges when the full correlation case would not exhibit it, {\em i.e}, panels (a)-(d). The results do not depend qualitatively on the value of $\gamma$. 
In order to characterize the range of values where partial degree-frequency correlation leads to explosive synchronization, Figure \ref{hist_area} 
shows the area $A$ between the forward and backward continuations of the synchronization diagrams $r(\lambda)$ for different values of the fraction $f$ of vertices for which degree-frequency correlation holds. The remaining vertices have natural frequencies draw from a power law distribution   with $\gamma = 3.0$ and $\omega_0=3$. For values of the parameter $\alpha > 0.3$, leading to network topologies with very mild heterogeneities, partial correlation does not promote ES and a second order phase transition is observed always. However, for $\alpha \le 0.3$, some realizations of networks and natural frequencies start to show a small hysteresis loop. As the parameter $\alpha$ decreases further, the frequency of realizations that show ES, as well as the area of the hysteresis loop, start both to grow.

For the cases shown in Figure \ref{hist_area}, the optimal fraction of vertices $f$ that must be correlated in order to maximize the hysteresis area seems to be around $f = 0.1$ and are roughly independent of the parameter $\alpha$. However, for $\alpha = 0.0$, that corresponds to a scale free network, the average area of the hysteresis loop attains higher values than when full degree-frequency correlation holds.

\subsection{A benchmark biological network}

We performed also some numerical experiments with a real biological network, namely the neural network of the worm {\em Caenorhabditis elegans} \cite{watts1998}. We note that we do not claim that the Kuramoto model and explosive synchronization play any role in the biology of the neural system of the worm {\em C. Elegans}. We use this network only as an example of a real world network \cite{skardal2014}.

\begin{figure}
\centering
\includegraphics[scale=0.20]{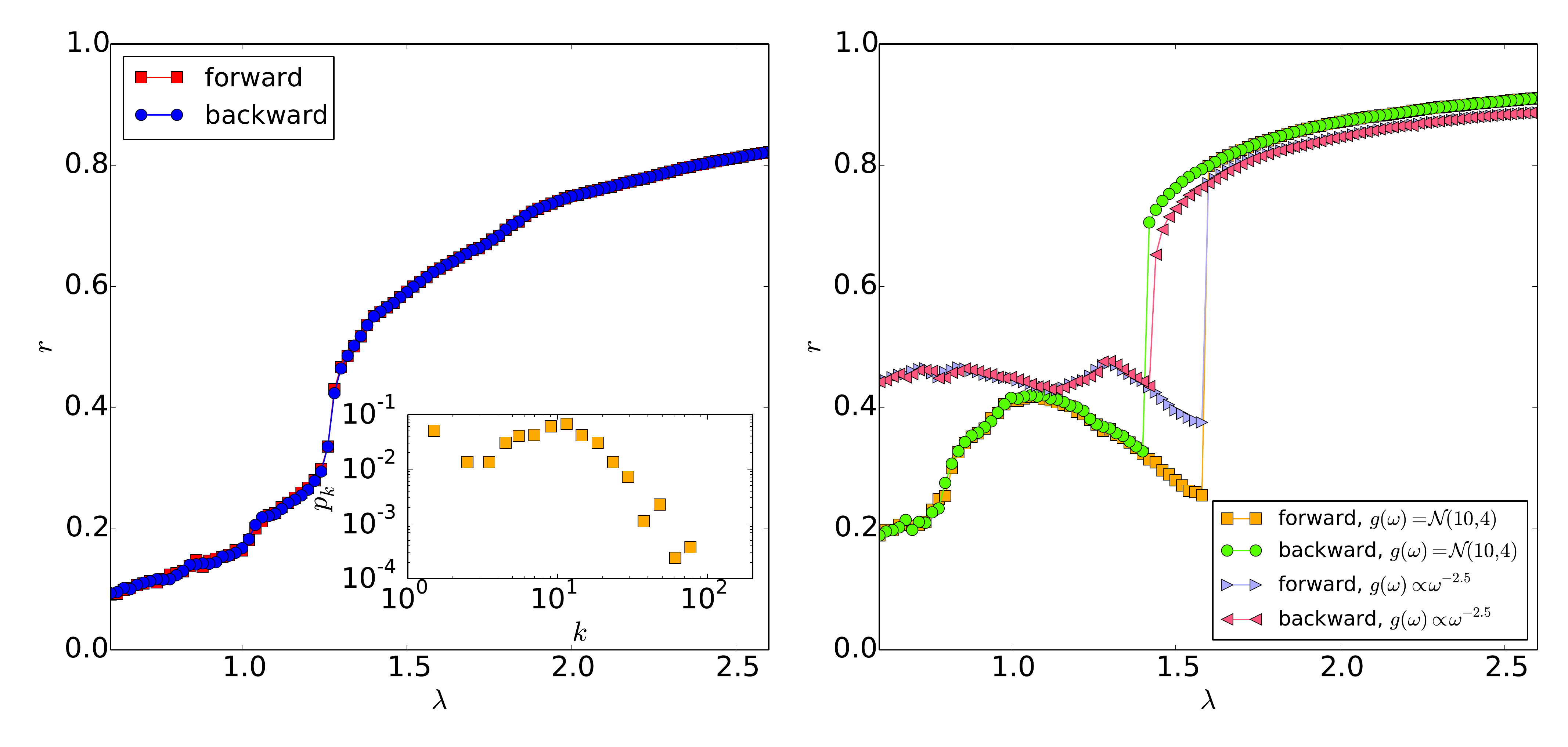}
\caption{The graphics show the synchronization diagrams $r(\lambda)$ for the neural network of the worm {\em C. elegans}. The diagram on the left was calculated assuming $\omega_i = k_i$ for every vertex $i$ of the network. The inset shows the degree distribution of the network. The diagram at right corresponds to a partial degree-frequency correlation: the correlation holds only for the 20 vertices with largest degrees. The natural frequencies for the remaining vertices were drawn from two different distributions, either from a power law distribution $g(\omega) \propto \omega^{-\gamma}$, with $\gamma  = 2.5$ and $\omega_0=3$ according to (\ref{powerlaw_definition}), or from a normal distribution with $\langle \omega \rangle = 10$ and $\sigma_0= 4$, see the insets. For the case of a power law, explosive synchronization holds also for other values of $\gamma$ (not shown).}
\label{celegans_diagram}
\end{figure}

We considered here the undirected and unweighted version of the network, which consists of $N = 297$ vertices representing the neurons of the worm and $M = 2148$ edges that roughly represent the synapses between the neurons. The graphics in the left side of the Figure \ref{celegans_diagram} depicts the diagram $r(\lambda)$ for the case of full degree-frequency correlation, again according to (\ref{corr}). We observe clearly a smooth second order phase transition, in agreement with previous works \cite{skardal2014}. However, when the degree-frequency correlation holds only for the 20 vertices with largest degree ($7\%$ of all vertices), whereas for the other oscillators their natural frequencies are drawn from either a power law or a normal distributions, we observe, remarkably, a very pronounced first order, explosive, transition with the typical hysteresis loop. Note that the hysteresis loop is present for two distributions with very different characteristics, sugesting that explosive synchronization may be seen for a very large range of parameters.

\begin{figure}
\centering
\includegraphics[scale=0.20]{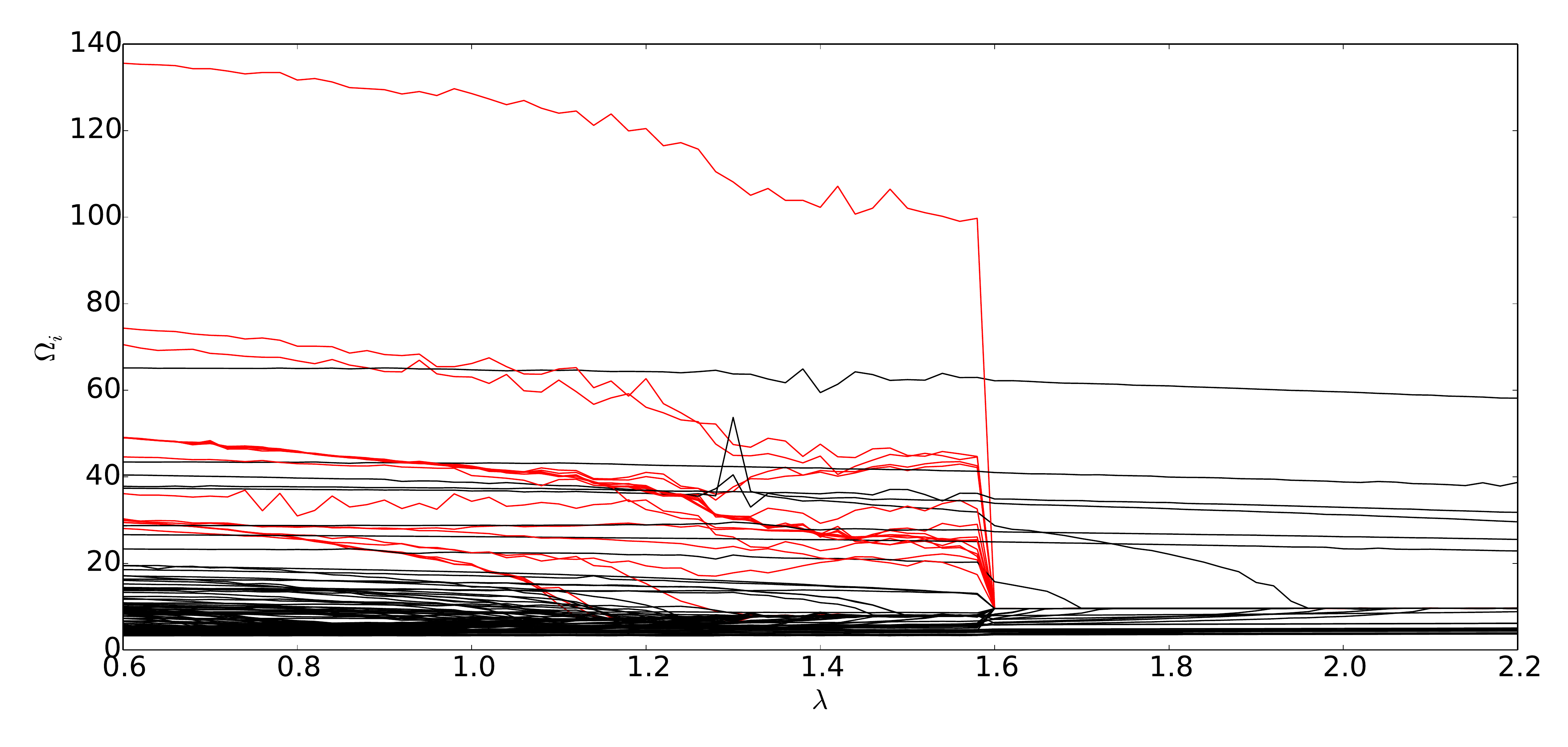}
\caption{The Figure shows the evolution of the effective frequencies $\Omega_i$ of the oscillators on the forward continuation for the {\em C. elegans} neural network, for the case with $\gamma = 2.5$ of Figure \ref{celegans_diagram}. The red lines correspond to the vertices for which degree-frequency correlation holds, whereas black lines represent the remaining ones.}
\label{celegans_frequencies}
\end{figure}

The effective frequencies (\ref{effective_freq}) of the oscillators on the forward continuation of Figure \ref{celegans_diagram} with $\gamma = 2.5$ are shown in Figure \ref{celegans_frequencies}. 
Above the critical coupling, almost all the oscillators, including all of which have frequency-degree correlation, collapse to a common frequency. However, at this point some oscillators (which correspond to less than $15\%$) still rotate with their own effective frequencies, only locking to the mean frequency at higher values of the coupling.

\section{Final remarks}

Here, we have studied the existence of explosive synchronization in Kuramoto models when the degree-frequency correlation holds only for a small set of the vertices of the network. We have performed a mean-field analysis and calculated the critical coupling $\lambda_c$ corresponding to the onset of synchronization for several situations. 
We have found that when the correlations holds for the hubs, the vertices with the highest degrees, explosive synchronization not only still holds, but can also emerge  in situations which otherwise it would be absent, as seen in the panels (a) and (d) of Figure \ref{ba_er_diagram}, as well as in the case of the neural network of {\em C. elegans}, Figure \ref{celegans_diagram}. 

We use our numerical simulations to go further the mean-field analysis. In particular, we found that partial degree-frequency correlation results in discontinuous synchronization transitions in a large range of parameters, caracterizing both the network and the natural frequency distribution, as we found ES in networks with only mild heterogeneities (parameters $\alpha \ge 1$ in Figure \ref{hist_area}) or in real world networks as in the case of the {\em C. elegans} neural network. With respect to the natural frequencies of the non-correlated oscillators, ES is observed for different values of $\gamma$ in Figure \ref{ba_er_diagram} and even for Gaussian distributions in the case of Figure \ref{celegans_diagram}.
\begin{figure}
\centering
\includegraphics[scale=0.20]{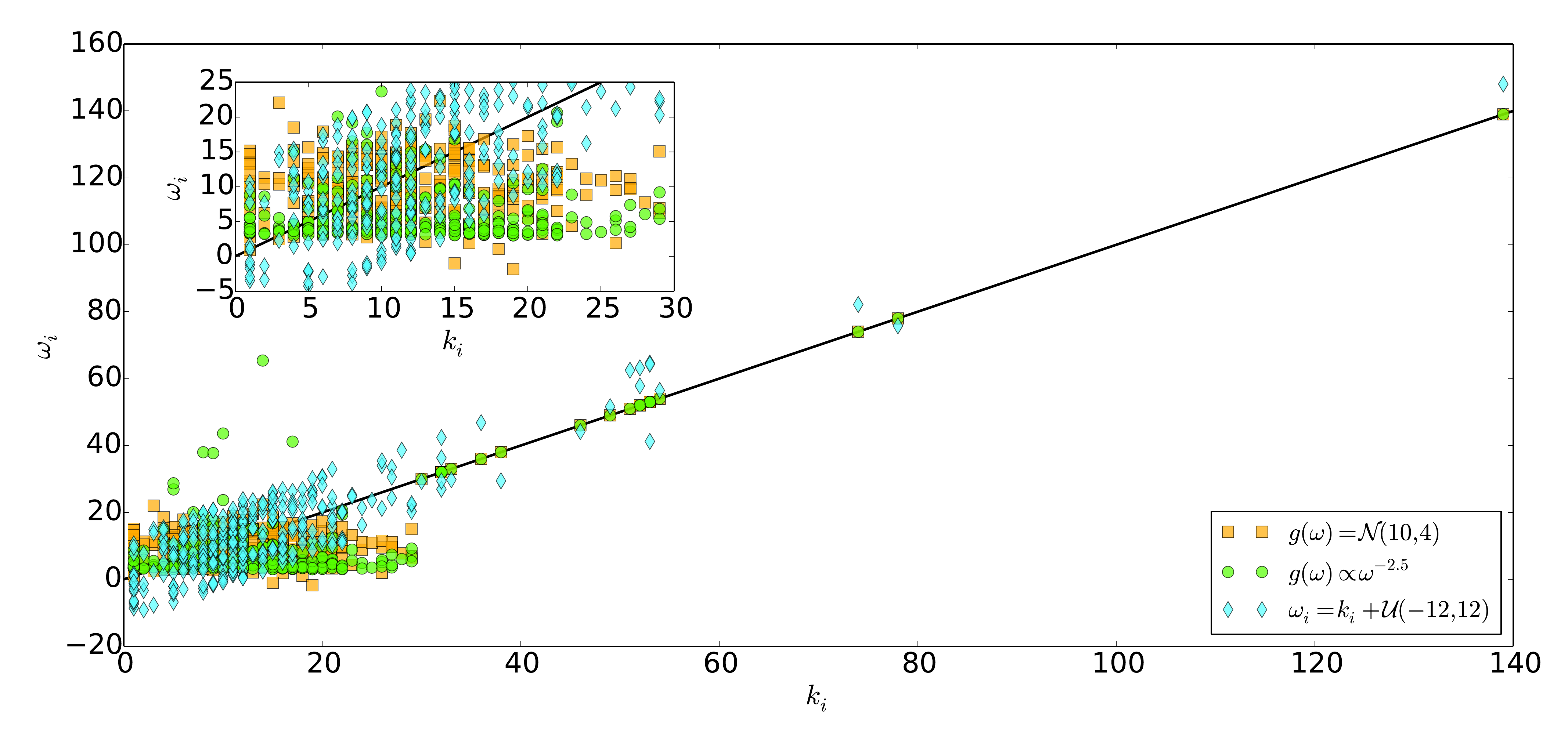}
\caption{The Figure shows the natural frequencies $\omega_i$ of the oscillators as a function of their degrees $k_i$ for the cases analysed in Figure \ref{celegans_diagram}. Three different cases are depicted, when there is partial correlation, with the non-correlated vertices having a natural frequency draw from either a power law or a normal distribution as well as when there is a quenched disorder, $\omega_i = k_i + \zeta_i$, where $\zeta_i$ is a random variable drawn from the uniform distribution $\mathcal{U}(-\epsilon, \epsilon)$, with $\epsilon = 12$. The inset shows the region for small degrees and frequencies.}
\label{celegans_frquencies}
\end{figure}
It is also interesting to notice that comparing with the case of full correlation, the synchronization deteriorates when partial degree-frequency correlation holds. This can be seen from the smaller values of $r$ in the lower panels of Figure \ref{ba_er_diagram} as well as in Figure \ref{celegans_frequencies}, where drifting oscillators remain even after the threshold.

The problem of partial correlation was already analyzed in \cite{gomez-gardenes2011}, but for the case of random correlations. They showed that for a scale free network with exponent $\gamma = 2.4$, no ES is observed when less than around $50\%$ of the vertices were subjected to degree-frequency correlation. We   indeed confirm that for a Barabasi-Albert network with $\langle k \rangle = 6$ and $N = 400$ vertices, the threshold for ES is around $80\%$ when the correlated vertices are chosen randomly. On the other hand, when considering the hubs,
ES appears with only $10\%$ of the vertices subjected to degree-frequency correlation. One can understand qualitatively these results   by analyzing how synchronization is achieved in heterogeneous topologies. It is known that for scale free networks \cite{gomez-gardenes2007}, the synchronization emerges from a central core made by the hubs. As the coupling strength increases, this core recruits the poorly connected vertices to the synchronized cluster. With the degree-correlation for hubs, the frequency mismatch prevents as long as possible the central core of forming. However, when the central core forms, it has such a high value of $\lambda$ that a substantial fraction of vertices synchronize together.

Our results also agree with, and indeed expand, those ones presented in \cite{skardal2014}, where it is shown that when the correlation has a quenched disorder, $\omega_i = k_i + \zeta_i$, where $\zeta_i$ is a random variable uniformly drawn from the range $(-\epsilon, \epsilon)$, explosive synchronization is still observed and, moreover, it can be seen in networks such as the {\em C. elegans} neural network. This happens mainly because, as we have shown here, the hubs have a key role in the synchronization process. The quenched disorder effectively uncorrelates the frequency and degree for small degree vertices, but the hubs, with their higher degrees, are still fairly correlated, even with the quite large values of $\epsilon$ values used in \cite{skardal2014}. We can see it from Figure \ref{celegans_frquencies}, where we show the natural frequencies $\omega_i$ as a function of the degree 
$k_i$ for the cases analysed in Figure \ref{celegans_diagram}, when the network has partial degree-frequency correlation, with the non-correlated vertices having a natural frequency draw either from a power law or from a normal distribution. The figure also show the case of quenched disorder of \cite{skardal2014}. For vertices with high degree, the natural frequencies for the three cases are all very similar, whereas in the region of small values of degree, the distributions of frequencies overlap over a considerably area for the three cases. These conclusions do not depend on the network details and could be indeed considered universal.

\begin{acknowledgments}
The authors thank CNPq, CAPES and FAPESP (grant 2013/09357-9) for the financial support and the anonymous referees for the very useful comments and suggestions. AS thanks Prof. Leon Brenig for several discussions and for the warm hospitality at the Free University of Brussels, where the initial part of this work was done. 
\end{acknowledgments}

\appendix

\section*{Appendix}
We compile in this Appendix  the evaluation of the pertinent integrals of the mean-field analysis of the Section 2.
We start with the simpler integral  $I_2(\lambda r)$ given by
\begin{equation}
\lambda r I_2(\lambda r) = \int_{k_{\rm min}}^{\infty} dk\, k
\int_{\Omega-\lambda kr}^{\Omega+\lambda kr} d\omega \,  G(\omega,k) 
    \exp\left(i \arcsin \frac{\omega - \Omega}{\lambda kr} \right). 
\end{equation}
By introducing the new variable $\omega = \Omega + \lambda k r \eta$, one has
\begin{equation}
\label{I_2}
  I_2(\lambda r) =  
\int_{-1}^{1} d\eta \, 
    \exp\left(i \arcsin \eta \right) 
\int_{k_{\rm min}}^{\infty} dk\, k^2  G(\Omega + \lambda k r \eta,k)  .
\end{equation}
Let us now perform the integration in $k$ taking into account 
the definition (\ref{defG}) of $G(\omega,k)$. We get  
\begin{widetext}
\begin{equation}
\label{intk}
\int_{k_{\rm min}}^{\infty}  k^2  G(\Omega + \lambda k r \eta,k)\, dk =
\frac{1}{|1-\lambda  r\eta|}  \left( \frac{\Omega}{1-\lambda  r\eta}\right)^2 P\left( \frac{\Omega}{1-\lambda  r\eta}\right)
H\left(  \frac{\Omega}{1-\lambda  r\eta} - k_*\right) +
\int_{k_{\rm min}}^{k_*}  k^2 
g(\Omega + \lambda k r \eta)P(k) \, dk.
\end{equation}
\end{widetext}
Since the integration interval in $\eta$ is bounded and the integrand is regular, one can commute the limit 
$r\to 0^+$
and the integration operations to obtain
easily (\ref{limit1}).

The evaluation of $I_1(\lambda r)$ given by 
\begin{widetext}
\begin{equation}
\label{I_1}
i I_1(\lambda r) = \frac{1}{2\pi} \int_{k_{\rm min}}^{\infty} dk\, k^2  \int_1^{\infty}d\eta \, \sqrt{\eta^2-1} 
\int_0^{2\pi} d\phi \,e^{i\phi}
\left(\frac{G(\Omega + \lambda k r \eta,k) }{\eta -\sin\phi}   +
\frac{G(\Omega - \lambda k r \eta,k) }{\eta +\sin\phi} \right),
\end{equation}
\end{widetext}
in the new variable $\omega = \Omega + \lambda k r \eta$ 
is quite   more intricate.
Notice that one can reduce the $\phi$-integration to an integral on the
complex plane to obtain
\begin{equation}
\int_0^{2\pi}\frac{e^{i\phi}d\phi}{\eta+\sin\phi} = -2\pi i\frac{|\eta| - \sqrt{\eta^2-1}}{\sqrt{\eta^2-1}},
\end{equation}
valid for $|\eta|\ge 1$, reducing $I_1(\lambda r)$ to
\begin{eqnarray}
I_1(\lambda r) &=&   \int_{k_{\rm min}}^{\infty} dk\, k^2  \\  
& &  \times \int_1^{\infty}d\eta \, f\left(\eta
\right)  \left( 
G(\Omega + \lambda k r \eta,k) - G(\Omega - \lambda k r \eta,k)
\right) \nonumber
\end{eqnarray}
where
\begin{equation}
f(\eta) =  \eta  - \sqrt{\eta^2-1}.
\end{equation}
Due to the definition (\ref{defG}) of $G(\omega,k)$, this integral can separated in
two parts
\begin{equation}
I_1(\lambda r) = I_1^a(\lambda r) + I_1^b(\lambda r)
\end{equation}
with
\begin{eqnarray}
\label{I_1^a}
I_1^a(\lambda r) &=&  
\int^{\infty}_{\max\left(k_*\frac{\Omega}{1-\lambda r}\right)}dk\, k P(k) 
\frac{1}{\lambda r}f\left(\frac{k-\Omega}{\lambda k r}\right)\\
& &- \int_{k_*}^{\max\left(k_*\frac{\Omega}{1+\lambda r}\right)}dk\, k P(k) 
\frac{1}{\lambda r}f\left(\frac{\Omega-k}{\lambda k r}\right)\nonumber ,
\end{eqnarray}
where we have already performed the integration in $\eta$, and 
\begin{eqnarray}
\label{I_1^b}
I_1^b(\lambda r) &=&  \int_{k_{\rm min}}^{k_*} dk\, k^2 P(k)  \\  
& &  \times \int_1^{\infty}d\eta \, f\left(\eta
\right)  \left( 
g(\Omega + \lambda k r \eta) - g(\Omega - \lambda k r \eta)
\right) \nonumber
\end{eqnarray}
Since we are interested mainly in the limit
$r\to0^+$ for both integrals, let us consider the approximation
\begin{equation}
\label{approx}
\frac{1}{\lambda r} f\left(\frac{|k-\Omega|}{\lambda k r}\right) \approx \frac{1}{2}\frac{k}{| k-\Omega|}
\end{equation}
first in (\ref{I_1^a}),
valid for $r\to 0^+$ and $k\ne \Omega$. Assuming $P(k)$ regular at $k=\Omega$, we have that (\ref{I_1^a}) can be approximated
in the limit $r\to 0^+$ by
\begin{equation}
\lim_{ r\to 0^+ }I_1^a(\lambda r) = {\rm p.v.\,} \left(\frac{1}{2}  \int_{k_*}^{\infty} dk\, \frac{k^2 P(k)}{k-\Omega}\right),
\end{equation}
where p.v. stands to the Cauchy principal value for the integral. Notice that a finite limit for this integral 
will typically require   that $\Omega \neq k_*$.

In order to evaluate $I_1^b(\lambda r)$ for $r\to 0^+$, let us restore the original variable $\omega$ in (\ref{I_1^b})
\begin{eqnarray}
\label{I_1^b1}
I_1^b(\lambda r) &=&  \int_{k_{\rm min}}^{k_*} dk\, k^2 P(k)\left[
\int_{\Omega+\lambda kr}^\infty d\omega \frac{g(\omega)}{\lambda kr} f\left( \frac{\omega-\Omega}{\lambda k r}\right) 
 \right. \nonumber \\   
& & \quad\quad\quad\quad- \left. 
\int^{\Omega-\lambda kr}_{-\infty} d\omega \frac{g(\omega)}{\lambda kr} f\left( \frac{\Omega-\omega}{\lambda k r}\right)
\right].
\end{eqnarray}
Since the integration interval in $k$ is bounded, one can take the limit $r\to 0^+$ directly. By using essentially the same
approximation (\ref{approx}), we have
\begin{equation}
\label{limit}
\lim_{r\to 0^+}I_1^b(\lambda r) = {\rm p.v.\,} \left(\frac{\beta}{2} \int_{-\infty}^{\infty} d\omega \frac{g(\omega)}{\omega - \Omega}\right),
\end{equation}
with $\beta$ given by (\ref{beta}).
Notice that, as expected, the limit (\ref{limit})  vanishes for symmetrical $g$ around $\Omega$, {\em i.e.}, for 
$g(\Omega +\omega) = g(\Omega - \omega)$. 

The evaluation of the principal value (\ref{pvgaussian}) for the standard Gaussian distribution involves the evaluation
of the principal value of the integral
\begin{equation}
I=
\int_{-\infty}^\infty \frac{e^{-s^2}}{s - s_0}\, ds,
\end{equation}
which one can calculate by  using the trick of differentiating under the integral sign. Notice that $I=g(1)$ with
\begin{equation}
g(x) = {\rm p.v.\,} \left(\int_{-\infty}^\infty \frac{e^{-x(s+s_0)^2}}{s  }\, ds\right)
\end{equation}
and that 
\begin{equation}
\label{de}
g'(x) = -s_0 \sqrt{\frac{\pi}{x}} - s_0^2 g(x),
\end{equation}
which is a linear differential equation for $g(x)$.  The solution of the homogeneous equation is simply  
$Ae^{-s_0^2x}$ and a particular solution can be obtained easily by setting 
$g(x) = e^{-s_0^2x}h(x)$, leading to the equation
\begin{equation}
h'(x) = -s_0\sqrt{\frac{\pi}{x}} e^{s_0^2x},
\end{equation}
which can be   integrated forwardly. 
The general solution for (\ref{de}) is
\begin{equation}
g(x) = e^{-s_0^2x}\left( A   - \pi  {\rm erfi}\left(s_0\sqrt{x}\right)\right),
\end{equation}
where 
\begin{equation}
{\rm erfi}(x) = \frac{2}{\sqrt{\pi}}\int_0^x e^{t^2}\, dt
\end{equation}
is the standard imaginary
error function. 
The integration constant $A$ can be determined from the requirement that $g(x)=0$ for $s_0=0$, leading to $A=0$. Taking $x=1$ one
gets
\begin{equation}
{\rm p.v.\,} \left(\int_{-\infty}^\infty \frac{e^{-s^2}}{s - s_0}\, ds \right)= - \pi e^{-s_0^2}  {\rm erfi}\left(s_0\right)
\end{equation}
and   (\ref{pvgaussian}) follows for the standard Gaussian distribution (\ref{gaussian}).

Finally, we have the evaluation of the integral
\begin{equation}
\label{inti}
A_\gamma = {\rm p.v.\,} \left( \int_{k_*}^{\infty}  \frac{k^{2-\gamma}}{k-\Omega}\, dk\right) ,
\end{equation}
which appears in the first term of (\ref{I_1}) for a power law degree distribution $P(k)\propto k^{-\gamma}$, with $\gamma >2$. Let us
consider first the case of integer $\gamma = n > 2$. Since 
\begin{equation}
\frac{1}{k^{n-2}(k-\Omega)} = \frac{1}{\Omega}\left(\frac{1}{k^{n-3}(k-\Omega)}-\frac{1}{ k^{n-2}}  \right),
\end{equation}
we have
\begin{equation}
A_n = \frac{1}{\Omega}\left(A_{n-1} - \frac{k_*^{3-n}}{n-3} \right),
\end{equation}
for $n>3$. This recurrence can be easily solved. Taking into account that 
\begin{equation}
A_3= {\Omega^{-1}}\log \left| \frac{k_*}{k_*-\Omega}\right|,
\end{equation} 
we have finally
\begin{equation}
A_n = {\Omega^{2-n}}\left(\log \left| \frac{k_*}{k_*-\Omega}\right| 
-  \sum_{\ell=1}^{n-3} 
\frac{1}{\ell} \left(\frac{\Omega}{k_*}\right)^{\ell} \right),
\end{equation}
valid for any integer $n>2$, from where (\ref{intn}) follows straightforwardly. The evaluation of (\ref{inti}) for non integer
values of $\gamma$
can be done by exploiting, for instance the series representation of $(k-\Omega)^{-1}$ for $k>\Omega$. We have
\begin{equation}
A_\gamma =  k_*^{2-\gamma} \sum_{\ell=0}^\infty\frac{\left(\frac{\Omega}{k_*}\right)^{\ell}}{\ell + \gamma -2 } ,
\end{equation}
valid for $k_*>\Omega$. For $0<k_*<\Omega$ we obtain analogously
\begin{equation}
A_\gamma = -k_*^{2-\gamma}  \sum_{\ell=0}^\infty
\frac{\left(\frac{k_*}{\Omega}\right)^{\ell+1}}{\ell - \gamma +3 }.
\end{equation}
Notice that the principal value of the integral
$
\int  d\omega \frac{g(\omega)}{\omega - \Omega}
$
for the case $g(\omega)\propto \omega^{-\gamma}$ employed in Section III can be evaluated analogously.

\end{document}